\documentclass[aps,preprint,showpacs,floats,epsf,epsfig,nofootinbib,12pt]{revtex4}
\usepackage{graphicx}
\usepackage{epsfig}
\usepackage[dvips]{color}

\def\be{\begin{eqnarray}}
\def\en{\end{eqnarray}}

\begin{document}

\title{An underlying symmetry determines all elements of CKM and PMNS up to a universal constant?}

\vspace{1cm}

\author{ Hong-Wei Ke$^1$\footnote{khw020056@hotmail.com.}and
        Xue-Qian Li$^2$\footnote{lixq@nankai.edu.cn}  }

\affiliation{  $^{1}$ School of Science, Tianjin University, Tianjin 300072, China \\
  $^{2}$ School of Physics, Nankai University, Tianjin 300071, China }

\vspace{12cm}

\begin{abstract}
Observing the CKM matrix elements written in different parametrization schemes, one can notice
obvious relations among the sine-values of the CP phases in those schemes.
Using the relations, we establish a few parametrization-independent equations, by which the matrix elements of the CKM matrix
can be completely fixed up to a universal parameter. If it is true, we expect that there should exist a hidden symmetry in the nature which determines
the relations. Moreover, it requires a universal parameter, naturally
it would be the famous Jarlskog invariant which is also parametrization independent. Thus the four
parameters (three mixing angles and one CP phase) of the CKM matrix are not free, but determined by the symmetry and the universal parameter.
As we generalize the rules to the PMNS matrix for neutrino mixing, the CP phase of the lepton sector is predicted
to be within a range of $0\sim 59^\circ$ centered at $39^\circ$  (in the P$_a$ parametrization) which will be tested in the future experiments.


\end{abstract}

\maketitle

\section{Introduction}
Since Cabibbo first noted the difference between the decays of
neutron ($n\to p+e^-+\bar\nu_e$) and muon ($\mu\to
e+\bar\nu_e+\nu_{\mu}$) and suggested there is a mixing angle
between d and s quarks\cite{Veltman},  and the mixing perfectly
explained the data of the $\beta-$decays of $\Sigma^-$ and
$\Lambda$ \cite{Cabibbo:1963yz}.   Later the CKM matrix
\cite{Kobayashi:1973fv} has been proposed to mix the three
generation quarks. In the $3\times 3$ matrix there are three
mixing angles and a CP phase which seem to be completely
independent of each other. The mixing is understood as that the
eigen-basis of weak interaction is not the same as that of mass,
so matching them, a unitary transformation matrix must be
introduced\cite{Fritzsch:1979zq,Li:1979zj,Fritzsch:1997fw}. From
then on, the research field about quark mixing has been thoroughly
investigated and exploration of its source has never ceased. In
analog to the quark sector, the Pontecorvo-Maki-Nakawaga-Sakata
(PMNS) matrix \cite{Maki:1962mu,Pontecorvo:1967fh} relates the
lepton flavor eigenstates with the mass eigenstates. Thus it is
natural to consider that there might be an underlying symmetry
which results in the practical CKM and PMNS matrices after
symmetry breaking. Recent studies on these matrices indicate that
there exist the quark-lepton complementarity and
self-complementarity\cite{Minakata:2004xt,Raidal:2004iw,Altarelli:2009kr,Zheng:2011uz,Zhang:2012pv,Zhang:2012zh,Haba:2012qx}
which hint an existence of a higher symmetry. All the progress in
this area inspires a trend of searching for whether such a hidden
symmetry indeed exists and moreover investigation of its
phenomenological implication is also needed.

The key point is to investigate whether there exist some relations
among the matrix elements of the CKM and/or PMNS matrices which
seem to be completely independent if there is no such a symmetry
to make an arrangement. There are nine parametrization schemes
which manifest the mixing in different ways. Therefore, we expect
that some relations among the parameters of these nine schemes
might hint a hidden symmetry if they indeed exist. Listing the
parameters (the mixing angles and CP phase) of the nine schemes in
a table and staring on them, we notice that there are relations
among the sine values of the CP phases in these schemes. Namely,
for the nine parametrization schemes, we find equalities among
those $\sin\delta_n$ i.e. $ \sin\delta_a\approx \sin\delta_e,\,
\sin\delta_b\approx \sin\delta_c,\, \sin\delta_d\approx
\sin\delta_e,\, \sin\delta_f\approx \sin\delta_h$ and $
\sin\delta_h\approx \sin\delta_i$ where the the subscripts $a$
through $i$ refer to the nine parametrization schemes. Considering
the relations not to be accidental, we would be tempted to believe
there is a hidden symmetry. Then associating { the expressions of
Jarlskog invariant which include sine values of the CP phases} in
all schemes with the experimentally measurable CKM matrix elements
$|U_{jk}|$ and using the above relations, we establish several
equalities which do not depend on any concrete parametrization
scheme at all. In principle when we apply the equalities to a
special scheme, the solutions of those equalities would give the
values of the parameters of the concerned scheme up to a universal
dimensionless constant.

What is the universal dimensionless constant? Surely it must be a scheme-independent quantity and has clear
physics significance. Naturally, one can conjecture that the Jarlskog invariant\cite{Jarlskog:1985ht} which is related to the CP violation
of hadrons is the best choice. Thus we would accept the allegation.

However, we also notice that all the equalities only approximately hold, even though the coincidence is very high. This can be well understood
that the hidden symmetry is slightly broken, but some characteristics of the original symmetry is partly retained after the breaking. The
approximate equalities are listed in the context of this work.

We generalize the relations among the CKM matrix to the PMNS case and find that
all the aforementioned equalities also hold for the lepton sector, even though the
accuracy is not as high as for the quark sector.  Along the line,
we further investigate the induced
phenomenological implication which may be tested in more accurate neutrino experiments.

The paper is organized as follows. After the introduction we
present those relations in section II. In section
III, we check these relations   numerically. In section IV we
will discuss the implications about the possible hidden symmetry and
draw our conclusion.

\section{Relations among elements of the CKM and PMNS matrices}

In this section we show how to obtain the relations among the elements of the CKM and as well the PMNS matrices.

\subsection{The mixing of fermions in standard model }

Mixing among different flavors of quarks (leptons) via the CKM (PMNS)
matrix has been firmly recognized and widely applied to phenomenological studies of hadronic processes.
The Lagrangian of the weak interaction
reads
\begin{eqnarray} \label{Lag}
\mathcal{L}=\frac{g}{\sqrt{2}}\bar {U}_L\gamma^\mu V_{CKM} D_L
W^+_{\mu}+\frac{g}{\sqrt{2}}\bar E_L \gamma^\mu V_{PMNS} N_L
W^+_{\mu}+h.c.,
\end{eqnarray}
where $U_L=(u_L, c_L, t_L)^T$,  $D_L=(d_L, s_L, b_L)^T$,
$E_L=(e_L, \mu_L, \tau_L)^T$and  $N_L=(\nu_1, \nu_2, \nu_3)^T$.
$V_{CKM}$ and $V_{PMNS}$ are the CKM  and PMNS matrices
respectively.  The $3\times 3$ mixing matrices are written as
    \begin{equation}\label{M1}
      V=\left(\begin{array}{ccc}
        U_{11} &U_{12} &U_{13} \\
         U_{21} &   U_{22} &  U_{23}\\
          U_{31} & U_{32} & U_{33}
      \end{array}\right).
  \end{equation}

Generally, for a $3\times 3$ unitary matrix
there are four independent parameters, namely three mixing angles
and one CP-phase. There can be various schemes to parameterize the
matrix and in literature, nine different schemes are presented and
widely applied. They are clearly listed in
Ref.\cite{Zhang:2012pv}. Here we try to collect those elements for
various parametrization schemes in Tab. \ref{tab1} where
$\theta_{nj}$ and $\delta_n$ are the mixing angles and CP-phase.
For clarity, we use $\vartheta_{nj}$ and $\delta'_n$ to denote the
corresponding quantities in the PMNS matrix.

\begin{table*}
      \caption{Nine different schemes for CKM matrix}\label{tab1}
      \begin{ruledtabular}
      \begin{tabular}{cccc}
      Scheme &  Jarlskog invariant& Mixing angles and CP-phase\\
      \hline
      ${{\rm P}_{a}:R_{23}(\theta_{a2})R_{31}(\theta_{a3},\delta_a)R_{12}(\theta_{a1})}$ & $J_a=s_{a1}s_{a2}s_{a3}c_{a1}c_{a2}c^2_{a3}\sin\delta_a$
      &
      $
        \theta_{a1}=\arcsin\frac{|U_{12}|}{|U_{11}|}, \theta_{a2}=\arctan\frac{|U_{23}|}{|U_{33}|}$\\&& $\theta_{a3}=\arcsin |U_{13}|,\delta_a=\left(69.10^{+2.02}_{-3.85}\right)^\circ
      $\\
      ${{\rm P}_{b}:R_{12}(\theta_{b3})R_{23}(\theta_{b2},\delta_b)R^{-1}_{12}(\theta_{b1})}$ & $J_b=s_{b1}s^2_{b2}s_{b3}c_{b1}c_{b2}c_{b3}\sin\delta_b$
       &
      $
        \theta_{b1}=\arctan\frac{|U_{31}|}{|U_{32}|}, \theta_{b2}=\arccos
        |U_{33}|$\\&&
       $ \theta_{b3}=\arctan\frac{|U_{13}|}{|U_{23}|},\delta_b=\left(89.69^{+2.29}_{-3.95}\right)^\circ
      $ \\
      ${{\rm P}_{c}:R_{23}(\theta_{c2})R_{12}(\theta_{c1},\delta_c)R^{-1}_{23}(\theta_{c3})}$ & $J_c=s^2_{c1}s_{c2}s_{c3}c_{c1}c_{c2}c_{c3}\sin\delta_c$
     &
      $
        \theta_{c1}=\arccos |U_{11}|,
        \theta_{c2}=\arctan\frac{|U_{31}|}{|U_{21}|}$\\&&
        $\theta_{c3}=\arctan\frac{|U_{13}|}{|U_{12}|}, \delta_c=\left(89.29^{+3.99}_{-2.33}\right)^\circ
     $ \\
      ${{\rm P}_{d}:R_{23}(\theta_{d2})R_{12}(\theta_{d1},\delta_d)R^{-1}_{31}(\theta_{3d})}$ & $J_d=s_{d1}s_{d2}s_{d3}c^2_{d1}c_{d2}c_{d3}\sin\delta_d$
      &
      $
        \theta_{d1}=\arcsin |U_{12}|, \theta_{d2}=\arctan\frac{|U_{32}|}{|U_{22}|}$\\&&$ \theta_{d3}=\arctan\frac{|U_{13}|}{|U_{11}|}, \delta_d=\left(111.95^{+3.82}_{-2.02}\right)^\circ
     $  \\
      ${{\rm P}_{e}:R_{31}(\theta_{e3})R_{23}(\theta_{e2},\delta_e)R^{-1}_{12}(\theta_{e1})}$ & $J_e=s_{e1}s_{e2}s_{e3}c_{e1}c^2_{e2}c_{e3}\sin\delta_e$
     &
      $
        \theta_{e1}=\arctan\frac{|U_{21}|}{|U_{22}|}, \theta_{e2}=\arcsin
        |U_{23}|$\\&&$
        \theta_{e3}=\arctan\frac{|U_{13}|}{|U_{33}|}, \delta_e=\left(110.94^{+3.85}_{-2.02}\right)^\circ
     $ \\
      ${{\rm P}_{f}:R_{12}(\theta_{f1})R_{31}(\theta_{f3},\delta_f)R^{-1}_{23}(\theta_{f2})}$ & $J_f=s_{f1}s_{f2}s_{f3}c_{f1}c_{f2}c^2_{f3}\sin\delta_f$
      &
      $
        \theta_{f1}=\arctan\frac{|U_{21}|}{|U_{11}|}, \theta_{f2}=\arctan\frac{|U_{32}|}{|U_{33}|}$\\&&$ \theta_{f3}=\arcsin
        |U_{31}|, \delta_f=\left(22.72^{+1.25}_{-1.18}\right)^\circ
      $\\
      ${{\rm P}_{g}:R_{31}(\theta_{g3})R_{12}(\theta_{g1},\delta_g)R^{-1}_{31}(\theta_{g2})}$ & $J_g=s^2_{g1}s_{g2}s_{g3}c_{g1}c_{g2}c_{g3}\sin\delta_9$
      &
      $
        \theta_{g1}=\arccos |U_{22}|,
        \theta_{g2}=\arctan\frac{|U_{23}|}{|U_{21}|}$\\&&$
        \theta_{g3}=\arctan\frac{|U_{32}|}{|U_{12}|}, \delta_g=\left(1.08^{+0.06}_{-0.06}\right)^\circ
     $\\
      ${{\rm P}_{h}:R_{12}(\theta_{h1})R_{23}(\theta_{h2},\delta_h)R^{-1}_{31}(\theta_{h3})}$ & $J_h=s_{h1}s_{h2}s_{h3}c_{h1}c^2_{h2}c_{h3}\sin\delta_8$
      &
      $
        \theta_{h1}=\arctan\frac{|U_{12}|}{|U_{22}|}, \theta_{h2}=\arcsin
        |U_{32}|$\\&&$
        \theta_{h3}=\arctan\frac{|U_{31}|}{|U_{33}|}, \delta_{h}=\left(157.31^{+1.18}_{-1.25}\right)^\circ
     $  \\
      ${{\rm P}_i:R_{31}(\theta_{i3})R_{12}(\theta_{i1},\delta_i)R^{-1}_{23}(\theta_{i2})}$ & $J_i=s_{i1}s_{i2}s_{i3}c^2_{i1}c_{i2}c_{i3}\sin\delta_i$
      &
      $
        \theta_{i1}=\arcsin |U_{21}|,
        \theta_{i2}=\arctan\frac{|U_{23}|}{|U_{22}|}$\\&&$
        \theta_{i3}=\arctan\frac{|U_{31}|}{|U_{11}|}, \delta_i=\left(158.32^{+1.13}_{-1.20}\right)^\circ
    $ \\
      \end{tabular}
      \end{ruledtabular}
\end{table*}

The P$_a$ parametrization i.e. P$1$ parametrization
in\cite{Zhang:2012pv}, can be realized via a serial rotations
 \begin{equation}\label{u1}
V=R_{23}(\theta_{a2})R_{31}(\theta_{a3},\delta_a)R_{12}(\theta_{a1}),
    \end{equation}
and the relevant rotation matrices $R_{23}, R_{31}$, $R_{12} $ were
given in Ref.\cite{Zhang:2012pv}.

Then for the P$_a$ parametrization an explicit expression is shown
as
 \begin{equation}\label{M1}
      V=\left(\begin{array}{ccc}
        c_{a1}c_{a3} & s_{a1}c_{a3} & s_{a3}\\
       -c_{a1}s_{a2}s_{a3} -s_{a1}c_{a2}e^{i \delta_a} & -s_{a1}s_{a2}s_{a3}+c_{a1}c_{a2}e^{i\delta_a} & s_{a2}c_{a3}\\
      -c_{a1}s_{a2}s_{a3} + s_{a1}s_{a2}e^{i\delta_a} & -s_{a1}s_{a2}s_{a3}-c_{a1}s_{a2}e^{i\delta_a} & c_{a2}c_{a3}
      \end{array}\right).
\end{equation}
Here $s_{aj}$ and $c_{aj}$ denote $\sin\theta_{aj}$ and
$\cos\theta_{aj}$ with $j=1,2,3$. The corresponding expressions in other schemes P$_n$ can be
obtained in similar ways.

Thanks to hard experimental measurements on the weak hadronic transitions
where the CKM matrix is involved, the mixing parameters for the
quark sector are \cite{Zhang:2012pv}
\begin{eqnarray} \label{qmix1}
\theta_{a1}=(13.023^{+0.038}_{-0.038})^\circ,
\theta_{a2}=(2.360^{+0.065}_{-0.038})^\circ,
\theta_{a3}=(0.201^{+0.010}_{-0.008})^\circ,
\delta_a=(69.10^{+2.02}_{-3.85})^\circ.
\end{eqnarray}

Similarly, the parameters in the  PMNS matrix \cite{Zhang:2012pv} are
\begin{eqnarray} \label{qmix2}
\vartheta_{a1}=(33.65^{+1.11}_{-1.00})^\circ,
\vartheta_{a2}=(38.41^{+1.40}_{-1.21})^\circ,
\vartheta_{a3}=(8.93^{+0.46}_{-0.48})^\circ,
\end{eqnarray}
which are directly measured by the neutrino-involved experiments,
especially the neutrino oscillations, but so far the CP-phase  $\delta'_a$ in the lepton sector
is undetermined yet.

\subsection{Several relations for CKM}

A close observation on the values of $\delta_n$ in different
schemes shows several approximate equalities
\begin{eqnarray} \label{rl1}
{\rm sin}\delta_a\approx {\rm sin}\delta_e,\, {\rm
sin}\delta_b\approx {\rm sin}\delta_c,\,  {\rm sin}\delta_d\approx
{\rm sin}\delta_e,\, {\rm sin}\delta_f\approx {\rm sin}\delta_h,\,
{\rm sin}\delta_h\approx {\rm sin}\delta_i.
\end{eqnarray}
It is well known that the Jarlskog invariant is independent of
different schemes, so using the above relations in Eq.(\ref{rl1})
and replacing the $s_{nj}$ and $c_{nj}$ with the ratios of modules
of corresponding elements {in the expressions of Jarlskog
invarian}, one can deduce several interesting relations among the
elements of CKM, which are fully experimentally measured values
and obviously parametrization-independent,
\begin{eqnarray} \label{rl2}
 \frac{\left({|U_{13}|}^2+{|U_{33}|}^2\right)
   \left({|U_{21}|}^2+{|U_{22}|}^2\right)}{{|U_{21}|} {|U_{22}|}
\left({|U_{23}|}^2-1\right)}\approx\frac{\left({|U_{11}|}^2+{|U_{12}|}^2\right)
    \left({|U_{23}|}^2+{|U_{33}|}^2\right)}{{|U_{11}|} {|U_{12}|}
    \left({|U_{13}|}^2-1\right)}\nonumber\\
   \frac{\left({|U_{12}|}^2+{|U_{13}|}^2\right)
   \left({|U_{21}|}^2+{|U_{31}|}^2\right)}{{|U_{11}|} \left({|U_{11}|}^2-1\right) {|U_{12}|} {|U_{21}|}}
  \approx\frac{\left({|U_{13}|}^2+{|U_{23}|}^2\right)
   \left({|U_{31}|}^2+{|U_{32}|}^2\right)}{{|U_{23}|} {|U_{32}|} {|U_{33}|} \left({|U_{33}|}^2-1\right)}\nonumber\\
 \frac{\left({|U_{13}|}^2+{|U_{33}|}^2\right)
   \left({|U_{21}|}^2+{|U_{22}|}^2\right)}{{|U_{21}|} {|U_{23}|} \left({|U_{23}|}^2-1\right) {|U_{33}|}}\approx\frac{\left({|U_{11}|}^2+{|U_{13}|}^2\right)
   \left({|U_{22}|}^2+{|U_{32}|}^2\right)}{{|U_{11}|} {|U_{12}|} \left({|U_{12}|}^2-1\right)
   {|U_{32}|}}\nonumber\\
 \frac{\left({|U_{12}|}^2+{|U_{22}|}^2\right)
   \left({|U_{31}|}^2+{|U_{33}|}^2\right)}{{|U_{12}|} {|U_{22}|} \left({|U_{32}|}^2-1\right)}\approx
   \frac{\left({|U_{11}|}^2+{|U_{21}|}^2\right)
   \left({|U_{32}|}^2+{|U_{33}|}^2\right)}{{|U_{11}|} {|U_{21}|} \left({|U_{31}|}^2-1\right)}\nonumber\\
 \frac{\left({|U_{12}|}^2+{|U_{22}|}^2\right)
   \left({|U_{31}|}^2+{|U_{33}|}^2\right)}{{|U_{12}|} {|U_{32}|} \left({|U_{32}|}^2-1\right) {|U_{33}|}}\approx
   \frac{\left({|U_{11}|}^2+{|U_{31}|}^2\right)
   \left({|U_{22}|}^2+{|U_{23}|}^2\right)}{{|U_{11}|} {|U_{21}|} \left({|U_{21}|}^2-1\right) {|U_{23}|}}.
\end{eqnarray}
Because the CKM is a unitary matrix the following relations should hold
\begin{eqnarray} \label{rl3}
{|U_{11}|}^2+{|U_{12}|}^2+{|U_{13}|}^2=1,
{|U_{21}|}^2+{|U_{22}|}^2+{|U_{23}|}^2=1,
{|U_{31}|}^2+{|U_{32}|}^2+{|U_{33}|}^2=1,\nonumber\\
{|U_{11}|}^2+{|U_{21}|}^2+{|U_{31}|}^2=1,
{|U_{12}|}^2+{|U_{22}|}^2+{|U_{32}|}^2=1,
{|U_{13}|}^2+{|U_{23}|}^2+{|U_{33}|}^2=1,
\end{eqnarray}
then Eq.(\ref{rl1}) can be simplified to forms of
\begin{eqnarray} \label{rl4}
 \frac{{|U_{21}|} {|U_{22}|}
}{
   1-{|U_{23}|}^2}\approx\frac{{|U_{11}|} {|U_{12}|}
   }{
    {|U_{23}|}^2+{|U_{33}|}^2}\nonumber\\
   \frac{{|U_{11}|}  {|U_{12}|} {|U_{21}|}}{
   1-{|U_{11}|}^2}
   \approx\frac{{|U_{23}|} {|U_{32}|} {|U_{33}|} }{
   1-{|U_{33}|}^2}\nonumber\\
 \frac{{|U_{21}|} {|U_{23}|}  {|U_{33}|}}{
   1-{|U_{23}|}^2}\approx\frac{{|U_{11}|} {|U_{12}|}
   {|U_{32}|}}{
   {|U_{22}|}^2+{|U_{32}|}^2}\nonumber\\
 \frac{{|U_{12}|} {|U_{22}|}}{
   1-{|U_{32}|}^2}\approx
   \frac{{|U_{11}|} {|U_{21}|} }{{|U_{11}|}^2+{|U_{21}|}^2
  }\nonumber\\
 \frac{{|U_{12}|} {|U_{32}|}  {|U_{33}|}}{{|U_{12}|}^2+{|U_{22}|}^2
   }\approx
   \frac{{|U_{11}|} {|U_{21}|}  {|U_{23}|}}{
   {1-|U_{21}|}^2}.
\end{eqnarray}

Since the denominators of all quantities on both sides of Eq.(\ref{rl2}) are small numbers and according to the
general rule for numerical computations, handling such large fractions  cannot guarantee higher accuracy,
thus we take their reciprocals to build Eq.(\ref{rl4}).

\subsection{The corresponding relations in PMNS}
Since the CP-phase in PMNS is not well determined so far, in
Ref.\cite{Zhang:2012pv} the authors tried to fix $\delta'_c$ in
the corresponding ${\rm P}_c$ parametrization and then used it to
determine $\delta'_n$ in other parametrizations. By the values
listed in Ref.\cite{Zhang:2012pv} we note that relations similar
to those in Eq.(\ref{rl1}) also exist. Thus we can write out the
relations in Eq.(\ref{rl4}), even though the new relations where
we replace $\delta_n$ in Eq.(\ref{rl1}) by $\delta'_n$, only
roughly hold. {Using the central value of $\vartheta_{a1}$,
$\vartheta_{a2}$ and $\vartheta_{a3}$ in Eq.(\ref{qmix2}) with
$\delta'_a=0^\circ$ we calculate the value in Eq.(\ref{rl4}) and
present the results in Tab. \ref{tab:value2}. It is noted that
those relations hold up to over 90\% accuracy in this case.}

{To solve the solar neutrino missing problem, it is suggested that
different flavor-neutrinos should also mix to compose physical
states (mass eigenstates). Later people learned that the mixing
patterns for neutrino may possess higher symmetries than for
quarks and among many other symmetric configurations, a favorable
form named as the  bimaximal mixing
pattern\cite{Vissani:1997pa,Baltz:1998ey,Barger:1998ta} is
proposed
\begin{equation}\label{BM}
|V_{TB}|=\left(\begin{array}{ccc}
     \frac{ 1 }{\sqrt{2}}& \frac{1}{\sqrt{2}} & 0\\
      \frac{1}{2} & \frac{1}{2} & \frac{1}{\sqrt{2}}\\
    \frac{1}{2} & \frac{1}{2} & \frac{1}{\sqrt{2}}
     \end{array}\right) \end{equation}
which manifests a high symmetry among the three flavors. Let us
investigate possible relations among those elements. Substituting
those elements $|V_{TB}|$  into Eq.(\ref{rl4}), one can find that
the equalities given above are exact, and it is a very interesting
issue. We will explore the exact solutions and their implications
in our future work.}
\section{Numerical  check}
To confirm the validity of the relations obtained in last section,
we should check them numerically. Using the central values of the
measured quantities:  $|U_{11}|$ = 0.97427, $|U_{12}|$ = 0.22535,
$|U_{13}|$ = 0.00352, $|U_{21}|$ = 0.2252, $|U_{22}|$ = 0.97344,
$|U_{23}|$ = 0.0412, $|U_{31}|$ = 0.00867, $|U_{32}|$ = 0.0404 and
$|U_{33}|$ = 0.999145\cite{PDG12} which are the best fit from
experiments, one can calculate all the concerned quantities in
Eq.(\ref{rl4}) (i.e. the combinations of the matrix elements).
{The} results are shown in the table \ref{tab:value}. It is not a
surprise to find that the first four equations hold with very high
accuracy whereas the last one declines from rigorous equality by a
few percents. Even though this deviation is not large, we still
feel puzzled about the reason, it might be caused by experimental
errors for measuring the involved quantities or there exists a
more profound cause. Indeed, in the last section, we will return
to make a short discussion on this issue.
\begin{table}
\caption{{The values in Eq.(10) for PMNS.}} \label{tab:value2}
\begin{tabular}{c|c|c|c}\hline
 Equality No. &Left side  &
 Right side& Relative deviation\\\hline
 1&0.49429   &0.461269 &6.68054\% \\
2&0.715365   &0.692966 & 3.1311\%  \\
3&0.392148   &0.375742  & 4.18372\% \\
4&0.49799  &0.449634  & 9.71014\% \\
5&0.376304  &0.35314 & 6.15571\% \\
\hline
\end{tabular}
\end{table}
\begin{table}
\caption{The values in Eq.(10) for CKM.} \label{tab:value}
\begin{tabular}{c|c|c|c}\hline
 Equality No. &Left side  &
 Right side& Relative deviation\\\hline
 1&0.219591   &0.219554 &0.0168495\% \\
2&0.973327   &0.972964 & 0.0372948\%  \\
3&0.00928607   &0.00934442  & 0.628361\% \\
4&0.219723  &0.219424  & 0.13608\% \\
5&0.00911122  &0.00952312 & 4.5208\% \\
\hline
\end{tabular}
\end{table}
Let us make more numerical computations.
If one uses equal signs to replace ``$\approx$" in Eq.(\ref{rl4})
which involve seven elements of $|U_{jk}|$, and takes $|U_{12}|$ =
0.22535, $|U_{23}|$ = 0.0412 and $|U_{33}|$ = 0.999145 as inputs,
by solving the four equations in Eq.(\ref{rl4}), he would obtain,
$|U_{11}|$= 0.974066, $|U_{21}|$= 0.225292, $|U_{22}|$ =0.972675
and $|U_{32}|$= 0.0401085 which are very close to the central
values of the data as shown above.

Furthermore, as we argued above, those equalities are scheme-independent, and there are only
four parameters for a unitary $3\times 3$ matrix, thus by solving the equations we may expect to get
all $|U_{jk}|$ up to a universal constant. Let us try and see if we can reach this goal.

As we guessed, the aforementioned equalities can determine all the
matrix elements up to a universal constant which could be the
scheme-independent Jarlskog invariant. Since the Jarlskog
invariant should be written in terms of the scheme-dependent
parameters (mixing angles $\theta_{nj}$ and CP phase $\delta_n$),
it is not convenient to use for our computations, instead, we
arbitrarily choose a few $|U_{jk}|$ values which are
experimentally measured, thus scheme-independent as inputs. By
manipulating the equations in  Eq.(\ref{rl3}), we have two
relations $U_{21}^2+U_{22}^2+U_{23}^2=1$ and
$U_{11}^2+U_{21}^2-U_{32}^2-U_{33}^2=0$. Using them and the first
three equalities in Eq.(\ref{rl4}) we obtain $|U_{11}|$= 0.974278,
$|U_{21}|$= 0.225159, $|U_{22}|$ =0.973451
 $|U_{32}|$= 0.0401402 and $|U_{33}|=$0.999151 by setting  $|U_{12}|$ =
0.22535 and $|U_{23}|$ = 0.0412. The results are satisfactorily consistent with the measured data. While
further including the fourth equalities in Eq.(\ref{rl4})
and only inputting one matrix element, say $|U_{12}|$, we find that directly solving the equation group
becomes very complicated and our computer program refuses to give a numerical result. Instead, as the iterative method is employed, with $|U_{12}|$ =
0.22535 as input, we
immediately obtain $|U_{11}|$=0.974278, $|U_{21}|$=0.225164, $|U_{22}|$ =0.973473
 $|U_{32}|$=0.039599, $|U_{32}|$=0.0406445 and $|U_{33}|=$0.999174. Comparing with the data, the predictions are even better than expectation.

\begin{center}
\begin{figure}[htb]
\begin{tabular}{cc}
\scalebox{0.7}{\includegraphics{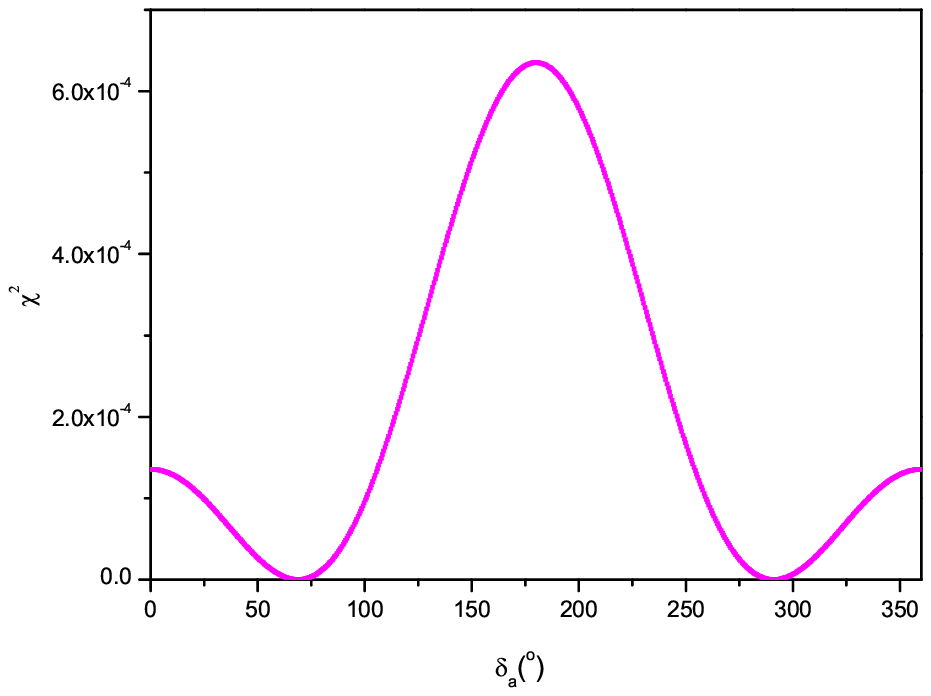}}&
{\scalebox{0.7}{\includegraphics{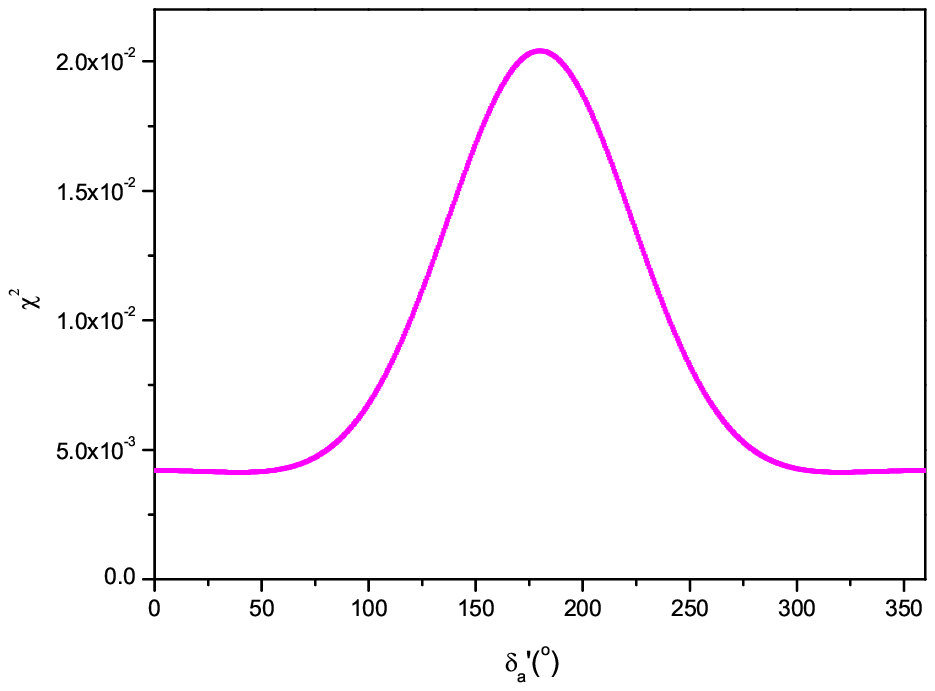}}}
\end{tabular}\\(a)\,\,\,\,\,\,\,\,\,\,\,\,\,\,\,\,\,\,\,\,\,\,\,\,\,\,\,\,\,\,\,\,\,\,\,\,\,\,\,\,\,\,\,\,\,\,\,\,\,\,\,\,\,\,\,\,\,\,\,\,
\,\,\,\,\,\,\,\,\,\,\,\,\,\,\,\,\,\,\,\,\,\,\,\,\,\,\,\,\,\,\,\,\,\,\,\,\,\,\,\,\,\,\,\,\,\,\,\,\,\,\,\,\,\,\,\,\,\,\,\,(b)
\caption{The dependence of $\chi^2$ which is defined in
Eq.(\ref{chi}) on the CP phase $\delta$ (a, for CKM ); and
$\delta'$ (b, for PMNS) .\label{fig:dep}}
\end{figure}
\end{center}

The four equations seem to be independent, so one may wish to fix
all the three  mixing angles and one CP-phase simultaneously.
However, it is obviously impossible because there are no any
numbers in those equations except 1. As all the $|U_{jk}|$'s are irrational
numbers, so one cannot expect to gain them by solving such equations. Instead,
we may try to find some valuable information about the CP phase
via the $\chi^2$ method\cite{Chiang:2004nm,Ke:2007ih}. For example
in P$_a$ parametrization the three CKM mixing angles are
$\theta_{a1}=13.023^\circ, \theta_{a2}=2.360^\circ,
\theta_{a3}=0.201^\circ$,  we would calculate $\chi^2$ which is
defined as
\begin{eqnarray}\label{chi}
\chi^2=(\frac{{|U_{21}|} {|U_{22}|} }{
   1-{|U_{23}|}^2}-\frac{{|U_{11}|} {|U_{12}|}
   }{
    {|U_{23}|}^2+{|U_{33}|}^2})^2+
   (\frac{{|U_{11}|}  {|U_{12}|} {|U_{21}|}}{
   1-{|U_{11}|}^2}
   -\frac{{|U_{23}|} {|U_{32}|} {|U_{33}|} }{
   1-{|U_{33}|}^2})^2+\nonumber\\
 (\frac{{|U_{21}|} {|U_{23}|}  {|U_{33}|}}{
   1-{|U_{23}|}^2}-\frac{{|U_{11}|} {|U_{12}|}
   {|U_{32}|}}{
   {|U_{22}|}^2+{|U_{32}|}^2})^2+
 (\frac{{|U_{12}|} {|U_{22}|}}{
   1-{|U_{23}|}^2}-
   \frac{{|U_{11}|} {|U_{21}|} }{{|U_{11}|}^2+{|U_{21}|}^2
  })^2
   \end{eqnarray}
The result is depicted in Fig. \ref{fig:dep}(a). One can notice
that there exist two minima of $\chi^2_{min}$ at $\delta_a\approx
68.75^\circ$ which fall in the tolerance range of the  data
$(69.10^{+2.02}_{-3.85})^\circ$ and its complementary position at
$291.06^\circ$.

Then we extend the same calculation to the PMNS case. With the
mixing angles $\vartheta_{a1}=33.65^\circ,
\vartheta_{a2}=38.41^\circ$ and $\vartheta_{a3}=8.93^\circ$ in
P$_a$ parametrization which are experimentally measured with
certain errors, we compute corresponding $\chi^2$ which depends on
$\delta'_a$ and the result is presented in Fig. \ref{fig:dep}(b).
It is also noted, the value of $\chi^2$ does not vary much within
a relatively wide range $0^\circ$ to $59^\circ$ centered at
$39^\circ$. Therefore, it is hard to firmly fix the CP phase at
the lepton sector with this $\chi^2$ method, and one needs to wait
for the more accurate long-baseline neutrino experiments in the
future to determine its value. Our estimate merely indicates its
possible range.

\section{SUMMARY AND DISCUSSIONS}

As is well known, the CKM and PMNS matrices emerge due to the
mismatch between the mass eigenstates and flavor eigenstates in
the quark and lepton sectors respectively. {The masses of
different generations are not the same and compose the hierarchy
problem}. That is one of the unknown parts in particle physics.
There must be {sort of} association between the mass hierarchy and
mixing matrix. In a recent paper Xing\cite{Xing-1} reviews the
situation and extends the discussion from the quark sector to the
lepton sector. Merlo also discusses the neutrino masses and mixing
from continuous symmetries\cite{Merlo}. Especially, the mass
hierarchy problem of neutrinos i.e. it is either normal or
inverted, is seriously considered and the JUNO experiment which
will operate in China in a few years, will determine it.

There  are still more mysteries for the mixing, for example, the
complementarity relations between {elements of CKM and PMNS}
matrices and the self-complementarities {among the matrix elements
of PMNS}
\cite{Minakata:2004xt,Raidal:2004iw,Altarelli:2009kr,Zheng:2011uz,Zhang:2012pv,Zhang:2012zh,Haba:2012qx}.
It was studied that the obvious symmetry in PMNS can be traced to
the possible symmetric textures for lepton
masses\cite{Wangbin,Liu:2013oxa}, thus whether a fundamental
symmetry really exists and what it is, is still an unsolved
problem in our theoretical prospect. Therefore, we are tempted to
investigate the CKM and PMNS matrices from a new angle and hope to
gain a better understanding of the possible underlying theories.

We note several relations among the CKM matrix
parameters of the nine parametrization schemes
i.e. ${\rm sin}\delta_a\approx {\rm sin}\delta_e,\, {\rm
sin}\delta_b\approx {\rm sin}\delta_c,\, {\rm sin}\delta_d\approx
{\rm sin}\delta_e,\, {\rm sin}\delta_f\approx {\rm sin}\delta_h,\,
{\rm sin}\delta_h\approx {\rm sin}\delta_i$, by which we establish several scheme-independent equalities. Those relations may
imply existence of something which is independent of any concrete parametrization.

The Jarlskog invariant is a physically measurable quantity, so
must be the same for all the schemes which include the sine values
of the CP phase $\delta_n$.

By the observation, it was suggested that the mixing matrices
originate from a higher symmetry\cite{Lam:2011ip}, and the nine
parametrization schemes are merely various representations of the
unique physical mixing matrix, therefore the relations among the
sine values of the CP phases of the nine schemes actually reveal
the inherent essence of the symmetry. We check these equalities
numerically and find the first four hold with very high precision.
Moreover, if the underlying symmetry really exists, the elements
of the CKM matrix which seem to be in an anarchy state at first
glimpse,  might be well arranged according to a certain rule. Thus
we would see whether the CKM matrix elements can be completely
determined from those equalities up to a universal constant, which
is the Jarlskog invariant.

On the other aspect,  most of the symmetries in
the nature are somewhat broken. Mostly, the breaking might not be
too large, so people still can trace back to the original
symmetry. For example, we proposed that breaking of the
complementarity and self-complementarity between CKM and PMNS
matrices is due to the involvement of a sterile
neutrino\cite{Ke:2014hxa}.
Here we can observe that the relations
of the CKM matrix elements hold with high precision, but an
approximation degree is also noted as $0.02\%\sim 0.63\%$.
Especially, the fifth relation listed in Eq.(\ref{rl4}) deviates
from equality by $4.52\%$.

It is naturally conjectured that the relations obtained for the
CKM matrix can also be applied to the  PMNS matrix. Because the
PMNS matrix elements are not well measured as that of the CKM
matrix, we can only expect that the relations among the PMNS
matrix elements only approximately hold. For the lepton case we
can also estimate $\delta'_a$ in terms of the $\chi^2$ method, the
resultant $\delta'_a\approx 39^\circ$. But as shown in the Fig.
\ref{fig:dep}(b), the value $\chi^2$ does not vary much from
$0^\circ$ to $59^\circ$. Therefore, we cannot firmly determine the
CP phase for the lepton sector as much as we do for the quark
sector from the relations yet.

Our observation on the relations indicates that a symmetry at higher scale does exist and
after running down to the low energy scale its essential characters
remain.  Therefore, even though it is slightly broken, we still can trace back to the original symmetry.
Further studies on the higher symmetry is definitely needed
to eventually understand these relations.

\section*{Acknowledgement}

This work is supported by the National Natural Science Foundation
of China (NNSFC) under the contract No. 11375128 and 11135009.

\appendix

\section{}


\end{document}